\documentclass[12pt]{article}
\usepackage{graphicx}
\newcommand{\be}{\begin{equation}}
\newcommand{\ee}{\end{equation}}
\newcommand{\ba}{\begin{eqnarray}}
\newcommand{\ea}{\end{eqnarray}}
\newcommand{\baa}{\begin{eqnarray*}}
\newcommand{\eaa}{\end{eqnarray*}}
\newcommand{\bb}{\begin{thebibliography}}
\newcommand{\eb}{\end{thebibliography}}

\begin{document}
\title{Long-range potential and the fine structure of the diffraction peak}

%!\classification{02.08.77
%<Replace this text with PACS numbers; choose from this list:
%                \texttt{http://www.aip..org/pacs/index.html}>
%!}
%!\keywords      {hadrons interactions, large distances, Coulomb-hadron interference}

%\author{O.V. Selyugin}{
%  address={BLTPh, JINR, Dubna, Russia}
%}

\author{O.V. Selyugin,   J.-R. Cudell \\
  BLTPh, JINR, Dubna, Russia \\
  IFPA, AGO Dept., University of Li\`ege, Belgium
} 
\maketitle

%\begin{abstract}
 
\vspace{0.5cm}

 {\small The possibility of oscillations in the differential elastic cross section of hadron
  scattering at small momentum transfer is studied.
  It is shown that string-like quark potentials at large distances can lead to
 such small oscillations, and
 an  analysis of the experimental data at small $|t|$ allows
 the determination of the parameters of
 the potential.}
%\end{abstract}

%\maketitle

\vspace{0.5cm}

    The AKM theorem \cite{akm}
   predicts that the differential elastic cross section will contain a structure
   periodic in the scale $q = \sqrt{|t|}$
  for $t \rightarrow 0$ \cite{gnsosc}.
  We study this effect through models for the hadron-hadron interaction
  at large distances.
This will lead to a new parametrisation of experimental data with a high
   confidence level in a wide energy region.
We consider many experimental data on
 nucleon-nucleon elastic scattering at
small momentum transfer in a very large energy interval from
$\sqrt{s} = 2$ GeV to 1800~GeV. We find that the oscillations
are present in most of the datasets.

The differential cross sections are given by the formula
\ba
d\sigma /dt &=& \pi \ \left\{F^2_C (t)+ \left[1 + \rho^{2} (s,t)\right] \ Im F^2_N(s,t)
                                                          \right.  \nonumber \\
  & &\left. \mp 2 \left[\rho (s,t) +\alpha \varphi(s,t) \right] \ F_C (t) Im
  F_N(s,t)\right\},   \label{ds2}
\ea
where $F_C$ is the Coulomb scattering amplitude, $F_N$ the hadronic amplitude,
$\rho(s,t)$ the cotangent of its phase, and $\varphi(s,t)$ the Coulomb-hadron
interference phase~\cite{selphase}.
 Here we neglect the hadron spin-flip
amplitudes and take into account all parts of the electromagnetic amplitudes.
The spin-non-flip amplitudes are written in the standard
form for small momentum transfer:
$F_{C} = \mp 2 \alpha G^{2}(t)/|t|$ ; $F^N (s,t)=h(s) \left[1+\rho(s,t)\right]
\exp(B(s)t/2)$, where
$\alpha$ is the fine-structure constant  and $G(t)$ the  proton
electromagnetic form factor squared.
This formula is used by experimentalists to extract $\rho$ from their data
to obtain the value of $\rho(s,t)$.
If an additional periodic amplitude has
a sizeable real part $ReF_{osc}(s,t)$, the oscillation in the differential
 cross sections will be proportional to
\ba
\Delta[d\sigma /dt]_{osc} \sim
    2 \ ReF_{osc}(s,t) \ [ReF^{C}(t)+\rho(s,t)ImF^h(s,t)].   \label{delosc}
\ea
The determination of $\rho(s,t)$ and $Re F_{osc}(s,t)$ then clearly
depend on each other.
\begin{figure}
 \includegraphics[width=0.5\textwidth]{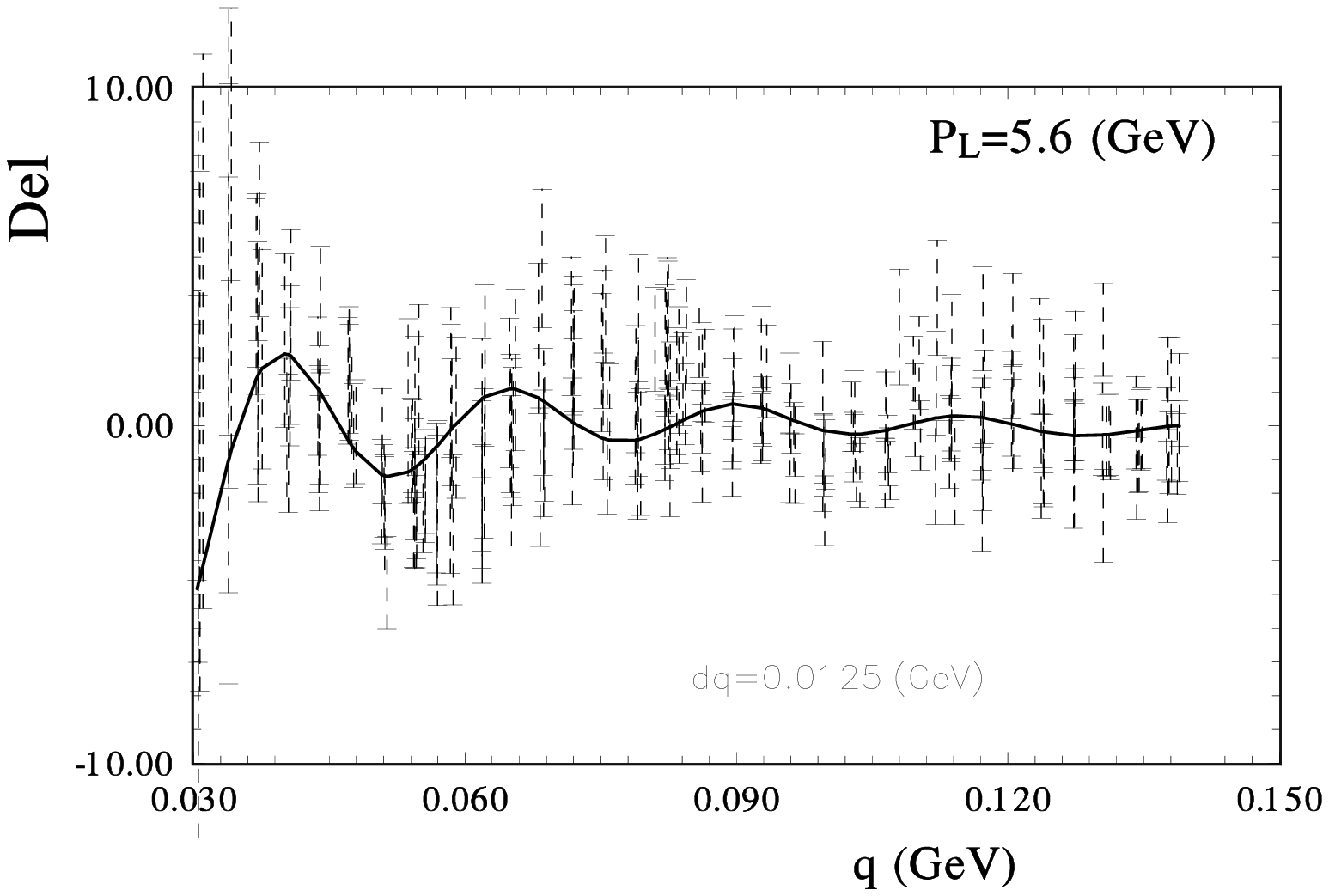}
%\vspace{-6.5cm}
 \includegraphics[width=0.5\textwidth]{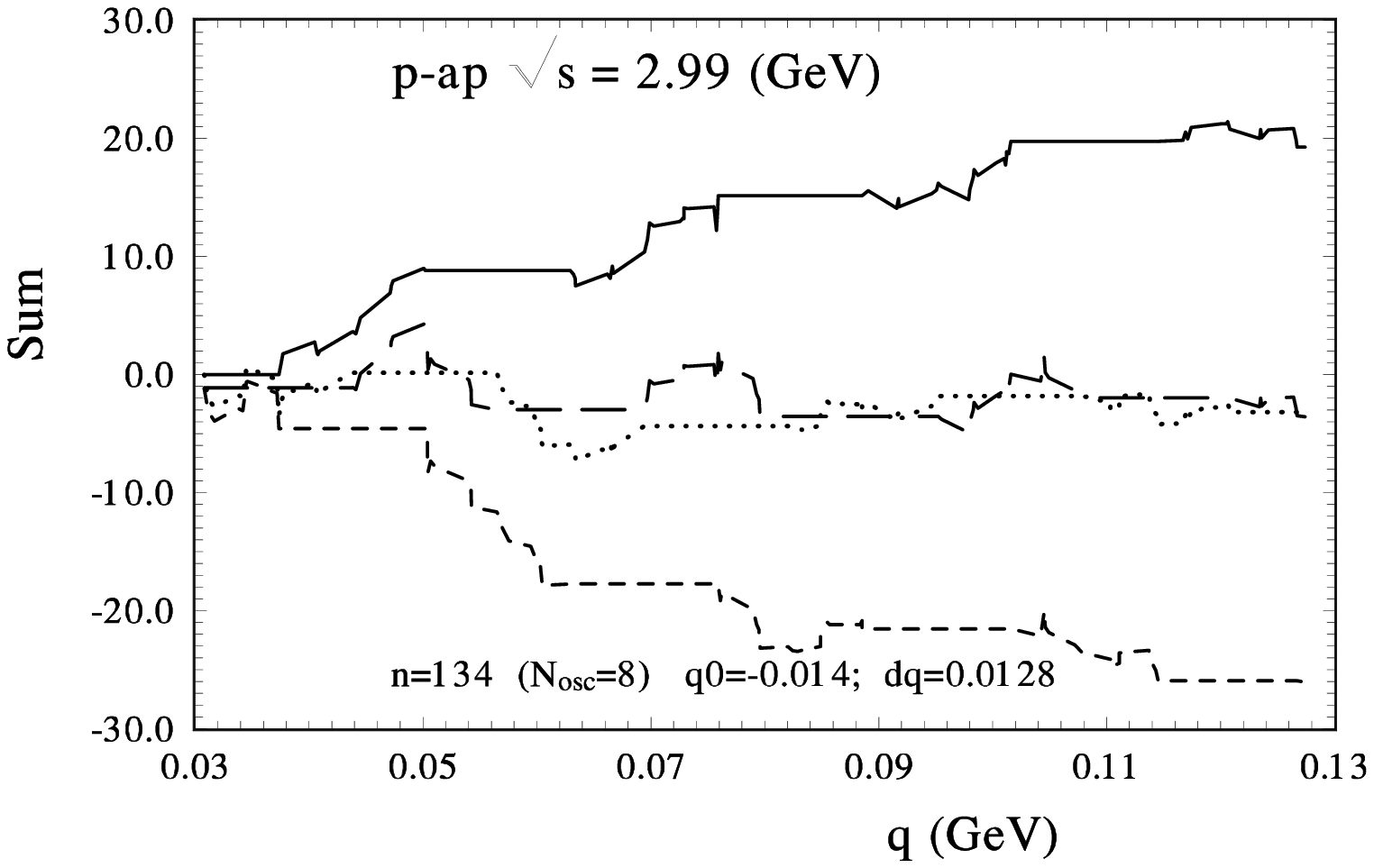}
\caption{ a) [left] Proton-antiproton scattering at $P_L =5.6$ GeV/c: points - the difference
between $d\sigma/dt$ and an exponential description; line - the fitted
additional function (\ref{oscill}) with $dq = 0.0125$ GeV/c.
b) [right] Proton-antiproton scattering at $P_L=3.7$ GeV/c:
the sum of the deviations of the experimental points from an exponential curve:
the solid and the short dashed lines are the sums on the even and odd intervals of width $dq = 0.0128$ GeV; the middle curves are the same but the beginning of the intervals is moved by $dq/2$.
  }
\label{Fig_1}
\end{figure}

We considered several forms for $F_{osc}$ (Bessel functions $J_0$ or $J_1$, sines, functions of $q$ or $q^2$...), and the best results are obtained for
   \be F_{osc}=h_{osc}\ \sin[\pi(\phi(s) + q/dq)].\label{oscill}\ee
We considered experimental data for $\bar p p$
 scattering at low energies ($\sqrt{s} = 3.1 - 6.2 $ GeV) and high energies
 ($\sqrt{s} = 52.6, \ 541,\ 546,\ 1800 $ GeV)  \cite{data-dis}.
 The inclusion of the term~(\ref{oscill}) in the fits leads to a decrease of the
 $\chi^2$ of the order of 20\%, as can clearly be seen in
Fig. 1a, where we show the deviation of the data from an exponential.
\begin{figure}
 \includegraphics[width=0.5\textwidth]{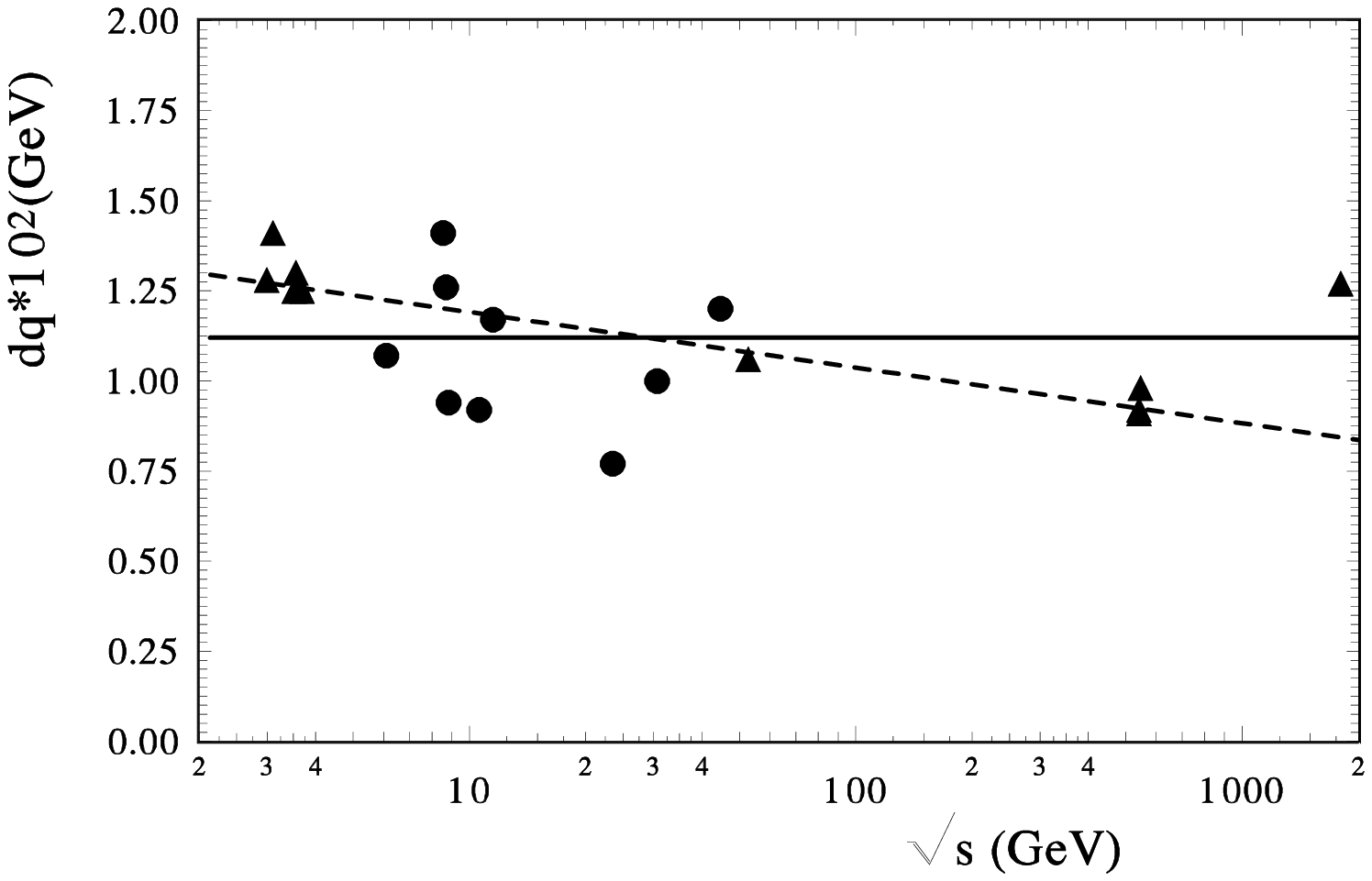}
%\vspace{-6.5cm}
 \includegraphics[width=0.5\textwidth]{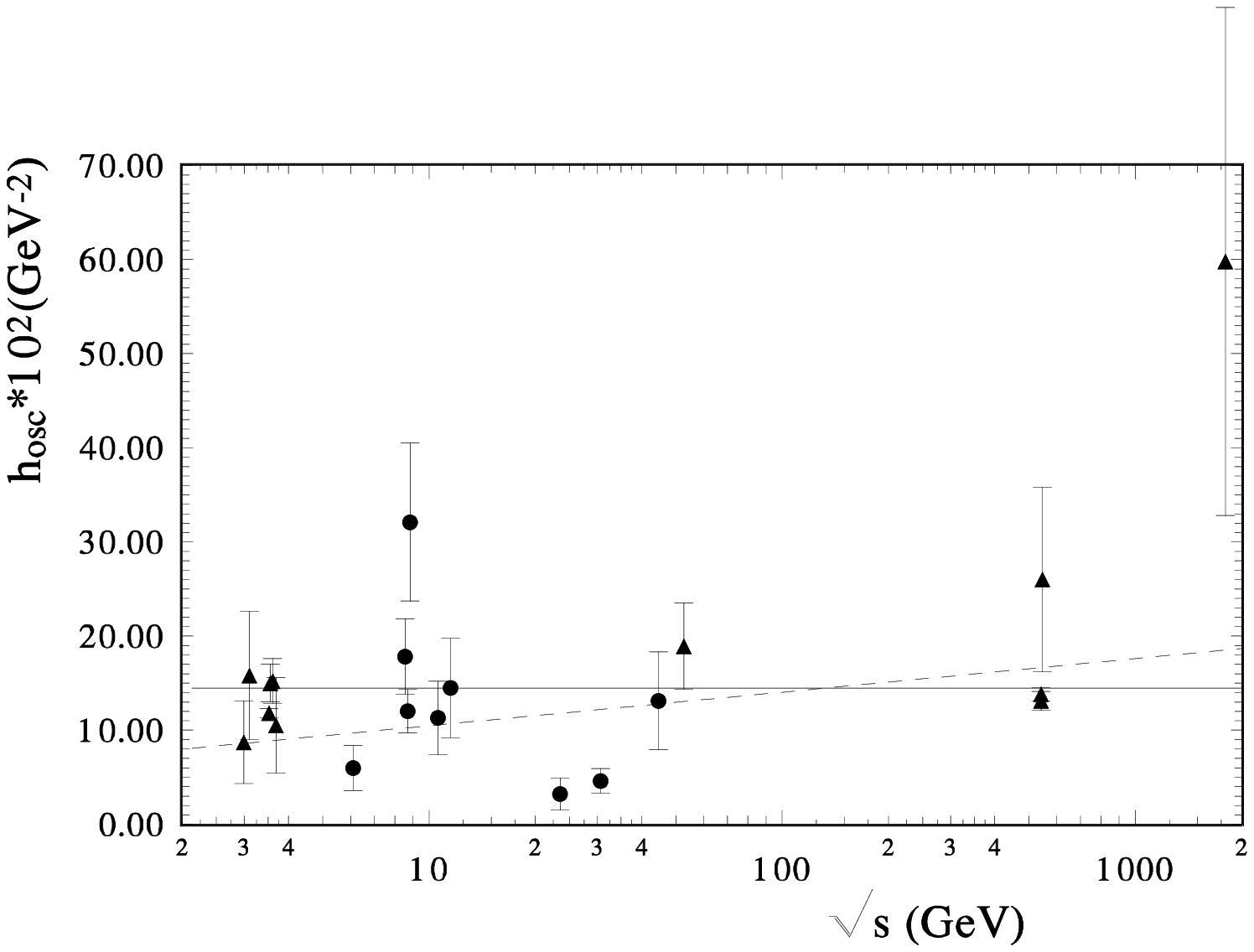}
\caption{ a) [left]
     Half-period of the oscillation;
b) [right]
Amplitude of the oscillation.
  }
\label{Fig_2}
\end{figure}
% \vspace {.5cm}
    For $pp$ elastic scattering we examined
   low energies  ($\sqrt{s} = 8.55,\ 8.7,\ 8.83,\ 10.6$ GeV)
   and high energies ($\sqrt{s} = 30.6,\ 44.7$ GeV).
   We again obtain an improvement in the $\chi^2$ between 15\% and 25\%.
More remarkably, the half period of the oscillation for all these experiments
lies near $12$~MeV (see Fig. 2a).  The normalization constant of the
 additional term fluctuates around $0.15$ GeV$^{-2}$, and it may
 grow with energy  as $\ln{s}$, as shown in Fig. 2b.

In all cases
 we obtain  a substantial decrease in the value of the $\chi^2$. It is clear that for
 such a high frequency, the improvement is unlikely to be accidental, or to
 correspond to  fluctuations of the data.
 We believe that it is
 evidence for the existence of such oscillations in the real part of the
 scattering amplitude.

 Note that to obtain the above fits, we also resorted to an unusual method.
 Indeed, the direct minimization of the $\chi^2$ works
 poorly, as one should first fix the model producing oscillations for the
 small-$|t|$ part of
 the scattering amplitude, and as the effect is small, so that outside
 of the exact fit, it will give an insignificant change to the $\chi^2$.

 Therefore, we also used a method comparing two statistically independent sets of data,
  for example \cite{hud}. The whole interval of $q$ is divided into small
  intervals $\Delta q_i,\ i=1,...n$,
  equal to $dq$. The deviations of the experimental data from an exponential,
  weighted by the inverse the experimental errors, are added in two separate sums
  for $i$ even or odd.
  If the intervals correspond to the period of the cross section, then
 these two sums
  will be significantly different in sign. If the starting point of the first interval
  is moved by one half period, the two sums will become identical, and close to zero
  (see
  the middle lines of Fig.~1b and \cite{J-LHC}).  We can calculate the significance
  level by comparing the two fits, with the interval moved or not.
   Most remarkably this method leads to a similar value for the half period,
   $dq \simeq 12$~MeV. Note that the effect comes mostly from the lowest $q$, where it
   gets enhanced by the Coulomb-hadron interference term.

  The convergence of the two methods to equivalent fits, and the significant improvement
  in the quality of the fit
  for many independent experiments, suggest that the oscillations are not due to
   statistical fluctuations, but rather reflect a physical phenomenon.

\section{The hadron potential at large distance}
Such a small period of oscillation may be related with the properties of
the hadron interaction at large distances.

In some calculations of the scattering amplitude, determined by the gravitational
potential in an $n+4$ dimensional world
in the framework of the ADD-scenario \cite{add},
 one obtains a periodical
structure (our calculations  and \cite{aref}).

In general, we can assume that an additional potential has a small constant value
at large distances and is sharply
     screened at a given distance, which provides a cut-off for the integrals.
   In the $q$-representation, the corresponding amplitude will oscillate
   with a period that depends on the distance at which
    the potential is cut, as shown in Fig. 3a.

    Such a scenario could also be realized during fireball processes
    via the screening of the electromagnetic interactions at large
    distances.

   Indeed, let us take the additional potential in our case in $b$-space
 in the following form:
\begin{equation}
F_{ad}(s,t) \sim
\int_{0}^{\infty} b db\ J_0(b q)\ h_{ad}(s)  \  [b_{scr}^2 - b^2]^{-2}  \ = \         iq b_{scr} K_1(i q b_{scr}),
\end{equation}
  where $b_{scr}$ is the distance at which the additional potential has a
screening effect, and   $K_1(ix)$
is the MacDonald function of imaginary arguments.
  If $b_{scr}$ is sufficiently large we obtain an oscillating amplitude with a small period in $q$ (see Fig. 3b). Note that in this case the integration has no specific cut.

However, the oscillations can be characterized by an amplitude and a period that do not
depend on energy, or depend on it very weakly.
This, together with the small size of the coupling and the long range of the
interaction, may point to an  electromagnetic origin of this effect.

\begin{figure}
 \includegraphics[width=0.4\textwidth]{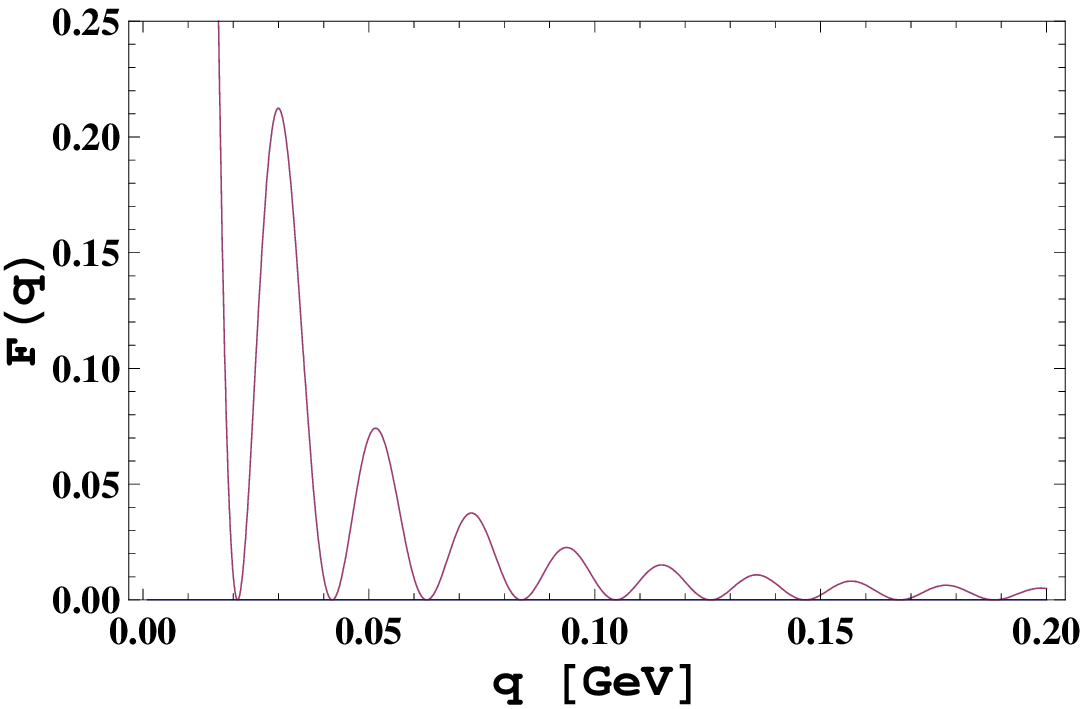}
\vspace{-0.5cm}
 \includegraphics[width=0.4\textwidth]{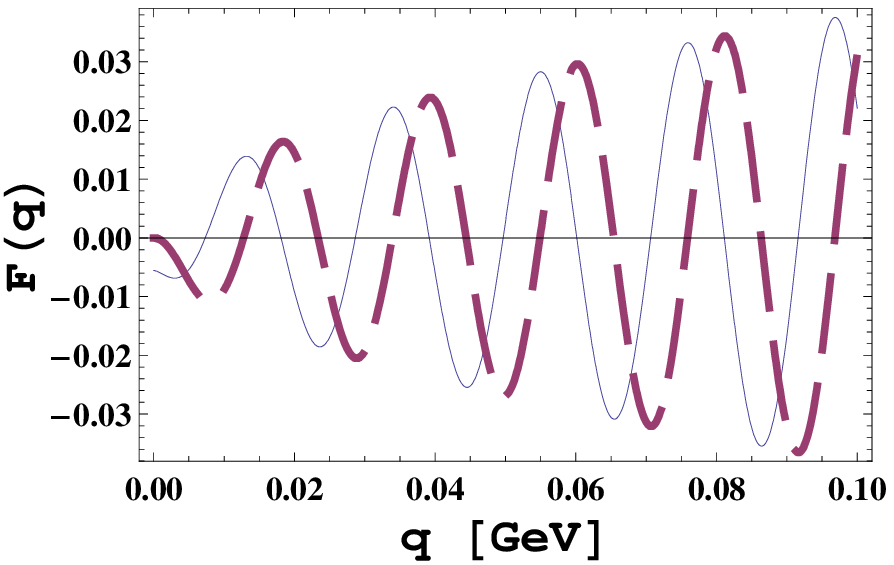}
\caption{ a) [left]
The amplitude corresponds to a small constant potential cut at large distance;
  b) [right]
 The additional amplitude obtained in the model of the rigid string
 (solid and dashed lines - the real and imaginary parts).
  }
\label{Fig_4}
\end{figure}

\section{Conclusion}
We have shown that oscillations, periodic in
$\sqrt{-t}$, exist in many experimental data sets at a significant level.
      The confirmation of the existence of such a periodic structure
      in the elastic-scattering amplitude at the LHC would give us important
      information about
      the behavior of the hadron interaction potential at large distances
      which may be connected with the problem of confinement.
      \\

%\begin{theacknowledgments}
{\bf acknowledgments} \\
{O.S. would like to thank the organizers R. Fiore and A. Papa for the invitation
and support of his participation in conference.
 Work supported in part by a grant from FNRS.}
% RFFI (grant N 10-02-08349z).}
%\end{theacknowledgments}

\end{document}